\documentclass[10pt,letterpaper]{article}
\usepackage[top=0.85in,left=1.5in,footskip=0.75in,marginparwidth=2in]{geometry}

\usepackage[utf8]{inputenc}

\usepackage{cite}

\usepackage{nameref,hyperref}

\usepackage[right]{lineno}

\usepackage{microtype}
\DisableLigatures[f]{encoding = *, family = * }

\raggedright
\usepackage{changepage}

\usepackage[aboveskip=1pt,labelfont=bf,labelsep=period,singlelinecheck=off]{caption}

\usepackage{lastpage,fancyhdr,graphicx}
\usepackage{epstopdf}
\fancyheadoffset[L]{2.25in}
\fancyfootoffset[L]{2.25in}

\usepackage{color}

\definecolor{Gray}{gray}{.25}

\usepackage{graphicx}

\usepackage{sidecap}

\usepackage{booktabs}

\usepackage{amsmath}

\usepackage{wrapfig}
\usepackage[pscoord]{eso-pic}
\usepackage[fulladjust]{marginnote}
\reversemarginpar
\usepackage{xcolor, soul}
\sethlcolor{yellow}

\begin{document}

\begin{flushleft}
{\Large
\textbf\newline{Water Self-Dissociation is Insensitive to  Nanoscale Environments}}
\newline
\\
Solana Di Pino\textsuperscript{a},
Yamila A. Perez Sirkin\textsuperscript{a},
Uriel N. Morzan\textsuperscript{b},
Ver\'onica M. S\'anchez\textsuperscript{a},
Ali Hassanali\textsuperscript{b,1},
Damian A. Scherlis\textsuperscript{a,1},
\\
\bigskip
{\bf a.} {Departamento de Qu\'imica Inorg\'anica, Anal\'itica
y Qu\'imica F\'isica/INQUIMAE, Facultad de Ciencias Exactas
y Naturales, Universidad de Buenos Aires, Buenos Aires, Argentina}
\\
{\bf b.} {International Centre for Theoretical Physics, Trieste, Italy}
\\
\bigskip
{\bf $^1$} E-mail: ahassana@ictp.it or damian@qi.fcen.uba.ar.

\end{flushleft}


\begin{abstract}
Nanoconfinement effects on water dissociation and reactivity remain controversial, despite their importance to understand the aqueous chemistry at interfaces, pores, or
aerosols. The pKw in confined environments has been assessed from experiments and simulations in a few specific cases, leading to dissimilar conclusions. Here, with the use of carefully designed \textit{ab-initio} simulations, we demonstrate that the energetics of bulk water dissociation is conserved intact to unexpectedly small length-scales, down to aggregates of only a dozen molecules or pores of widths below 2 nm. The reason is that most of the free-energy involved in water autoionization comes from breaking the O-H covalent bond, which has a comparable barrier in the bulk liquid, in a small droplet of nanometer size, or in a nanopore in the absence of strong interfacial interactions. Thus, dissociation free-energy profiles in nanoscopic aggregates or in 2D slabs of 1 nm width reproduce
the behavior corresponding to the bulk liquid, regardless of whether the corresponding nanophase is delimited by a solid or a gas interface. The present work provides a definite and fundamental description of the mechanism and thermodynamics of water dissociation at different scales with broader implications on reactivity and self-ionization at the air-liquid interface.
\end{abstract}



\newcommand{\keywords}{
	Confinement \textbullet\
	Water chemistry \textbullet\ 
	Ab initio calculations \textbullet\ 
	Reaction mechanisms \textbullet\ 
	Molecular dynamics
}







\section*{Introduction}
\label{introduction}

Confinement deeply affects the physical chemistry of water. 
In cavities or pores of a few nanometers,
its freezing point can drop as much as 50$^\circ$C \cite{pccp_3_1185, pccp_10_6039, pccp_12_4124},
while other dynamic and thermodynamic properties such as
diffusivity \cite{chem-euj_10_5689, jpcc_112_12334, jpcc_116_7021,
jpcc_117_3330, jpcc_121_7533},
dielectric constant \cite{acs-nano_3_1279, 
prl_117_048001,  jcplett_10_6292, science_diel},
melting enthalpy \cite{pccp_8_3223, jpcb_115_14196},
or thermal expansion \cite{lang_25_5076, acs-appmatint_8_35621}, experience significant changes.
Remarkable progress has been made in the investigation of
water at the nanoscale regime: spectroscopic and calorimetric 
techniques
\cite{pccp_3_1185, chem-euj_10_5689,jpcc_112_12334,  jpcc_116_7021,  jpcc_121_7533, pccp_8_3223,
jpcl_5_174, nature-nano_12_267}
hand in  hand with molecular simulations
\cite{jpcc_117_3330,  acs-nano_3_1279, 
jcplett_10_6292, nature_414_188,  pnas_105_39, PhysRevLett.102.050603,
jpcc_116_1833,tuckerman-jpcl, gaigeot},
have been ingeniously applied to probe and characterize the
structure, dynamics, and phase transitions of water
in nanometric domains.

Notwithstanding these achievements,  the role of confinement on chemical reactivity
is barely understood, mostly because of the complexity associated with the determination
of equilibrium and kinetic constants in these conditions. 
In particular, the dissociation of water in porous media, tiny aerosols, biomolecular pockets,
membranes, and other nanospaces,
represents a fundamental problem of extreme relevance for experimentalists, which
remains unsettled. 
Studies in nanodroplets, reverse micelles, and cavities within supramolecular
assemblies, all strongly suggest that chemical kinetics is substantially affected
with respect to the bulk phase,
producing accelerations of up to 6 orders of magnitude in some cases
\cite{dwars,angw_55_2, jacs_141_10585, chemsci_10_2566, annrevphyschem71_31,
naturenano}.
This has prompted researchers to focus on confined chemistry in the search of
alternative routes for chemical synthesis \cite{dwars,angw_55_2, chemsci_10_2566, naturenano}, to
elucidate chemical transformation in the atmosphere \cite{acr_51_1229}, or to explain enzymatic
catalysis in living organisms \cite{nature_nanotech2016, biophysj}.
Whereas the mechanisms behind these increased rates are still a matter of debate,
the main causes that have been postulated
include the over-concentration of reagents, extreme
pH changes, or interfacial and entropic effects. Among the latter, the equilibrium constants of
products, reagents, and solvent, all become critical \cite{nl4008198}.
Understanding how water dissociation is affected in nanoenvironments
together with its role in aqueous reactivity is thus a key pending question.

Speculations based on indirect experimental evidence and theoretical conceptions have been made 
on a possible enhancement of the aqueous acidity in nanocavities \cite{science_220_365}.
In a recent NMR study, on the other hand,  a single H$_2$O molecule confined in the pocket of a C$_{60}$
derivative was shown to be less acidic than bulk water \cite{chemcomm_54_13686}.
In any case, the effect of pore radius and of the nature of the interface on the dissociation
constant of water continues to be a basic question still unanswered from a general perspective.
Hence, simulations become essential and have in fact contributed a number of clues
that shed light on this matter. 

First-principles simulations proved useful to calculate the pKw in
bulk \cite{sprik,parrinello,joutsuka}, and
were also applied to investigate the autoionization in confinement
in various settings,  from the interlayer spaces of minerals to carbon nanotubes
\cite{geocosmo_75_4978,prl_119_056002,jpcl}.
In particular, Mu\~noz-Santiburcio and Marx
computed the self-dissociation constant in slit pores of 1 nm delimited
by FeS layers, at high temperature and pressure \cite{prl_119_056002}.
They claimed that the free-energy barrier to dissociation ($\Delta G_d$) 
experienced a reduction of more than 15\%, 
explained in terms of an increase in the dielectric constant arising from confinement.

In the present study, we use  \textit{ab-initio} molecular simulations in combination 
with an appropriate sampling coordinate to track how the water dissociation barrier 
is affected by system size, hence providing a general answer to the puzzle on whether and 
by how much the self-dissociation of water changes as the dimensions of the liquid
phase approaches the nanometer.
We formulate and respond the question of how big the aqueous domain must be to 
preserve the energetics and the mechanism of bulk water dissociation. 
The answer is surprising: an aqueous cluster of just a dozen molecules 
exhibits a dissociative free-energy profile that cannot be distinguished 
from the one corresponding to the bulk phase. 
Size effects only start to tally in an aggregate that has half this number of molecules, 
or in pores of widths below 1.6 nm, 
where solvation of the hydronium and hydroxide ions are severely hindered. 
Our work rationalizes this outcome with the help of data-science tools, 
and manages to reconcile previous, apparently contradictory findings, offering a 
comprehensive picture of water dissociation at the nanoscale.
Beyond its obvious impact on chemistry under confinement,
the results of our simulations have broader implications on how the pKw changes near hydrophobic 
interfaces such as the surface of water or aqueous solutions.

\section*{Results and discussion}
\label{results_discussion}

\textbf{Dissociation barrier of nanoconfined water}

\vspace{0.2cm}

The dissociation free energy of water in the bulk phase was first computed from \textit{ab-initio} molecular
simulations in the late 90's by Trout and Parrinello \cite{parrinello},
employing density functional theory (DFT). These authors adopted the O-H distance
of the dissociating bond as the reaction coordinate.
While this constraint proved to be appropriate to induce bond breaking and to form the ion pair, 
when the O-H separation exceeds a certain length, it can not prevent the recombination 
of the hydroxide with a different proton through
Grotthuss diffusion \cite{sprik}, and therefore the application of this scheme
seems to be unable to produce
a well separated OH$^-$-H$_3$O$^+$  pair.
Shortly after, Sprik proposed  the proton coordination $n_H$ as an alternative 
reaction coordinate that, by avoiding recombination, may
facilitate the separation of the ions \cite{sprik}. This variable,
implemented in some of the studies cited in the introductory section
\cite{geocosmo_75_4978, prl_119_056002, jpcl},
counts the number of protons
surrounding a given oxygen atom. Each H atom contributes
to $n_H$ with a quantity that varies continuously between 0 and 1, depending on the O-H distance
(see Supporting Information).
Thus, the proton coordination assumes fractionary values and  can be used to introduce a bias potential
that enforces dissociation in a particular H$_2$O molecule as $n_H$ changes from 2 to 1.

Very recently, the study of water dissociation from \textit{ab-initio} simulations
was extended to two dimensional sampling by combining
the coordination number with the donor-acceptor distance \cite{joutsuka}. 
These calculations have contributed what is possibly
the most accurate thermodynamic landscape for bulk water dissociation available so far. 
In comparing 2D with 1D sampling, however, the author noticed that the latter,
when based on the coordination number, yields a sufficiently accurate profile for a model of 64
molecules.
In the light of this precedent, we perform here one-dimensional sampling
using the proton coordination, which allows us to afford the large number of computationally
intensive calculations arising from the
different model systems considered in this work. We emphasize that for the present investigation
the precise value of $\Delta G_d$ is not as relevant as it is its dependence
on system dimensions.
For bulk water our methodology reproduces the free-energy profiles
reported from previous molecular dynamics simulations using comparable
schemes \cite{sprik,parrinello, geocosmo_75_4978,prl_119_056002,jpcl}.
These approaches tend to underestimate the experimental dissociation free-energy by
a few kcal/mol, which is ascribable to the DFT functionals.
Importantly, however, our conclusions are not tied to the energy barriers resulting from the
calculations, dependent on the particular numerical scheme.
Instead, they rely on the variation of these barriers  as a function of system size,
ostensibly more robust. The fact that the main
outcome of the present analysis
is grounded on the changes in the dissociative barrier rather than on the barrier itself,
makes it pretty much independent of the DFT functional or electronic structure method chosen,
and is a major strength of this study.
In the Supporting Information we include results establishing
that the level of theory employed herein accurately captures these changes
in the reaction energies arising from alterations in the solvation structure,
which is the critical asset in the present context.
These benchmark calculations show that our quantum-mechanical approach, while
remaining computationally efficient to undertake the large number of calculations involved,
can reliably describe such effects.

\begin{figure}
	\begin{center}
    \includegraphics[width=8.6cm]{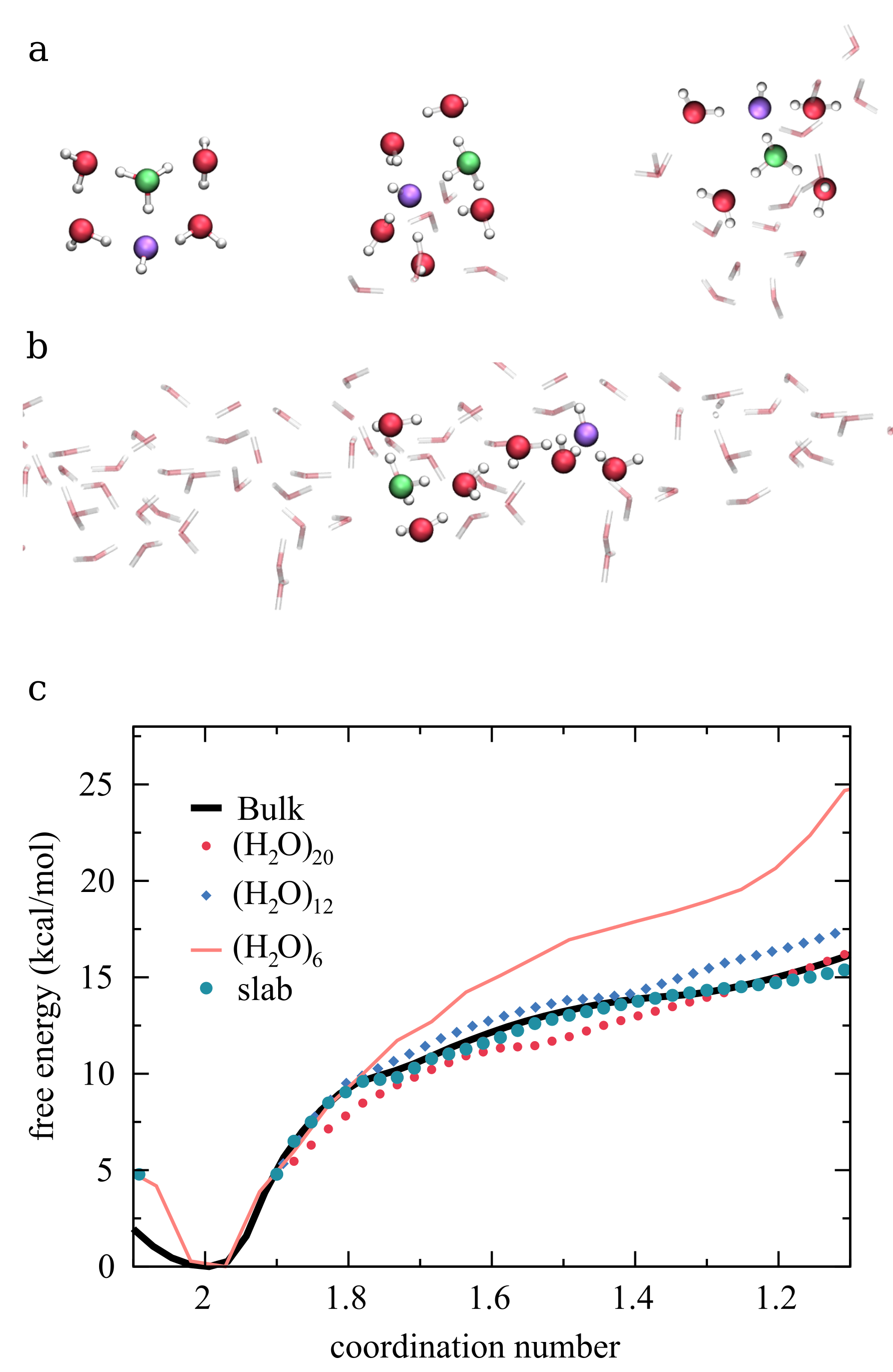}
		\caption{\textbf{a:} Snapshots of the (H$_2$O)$_6$, (H$_2$O)$_{12}$, and 
	(H$_2$O)$_{20}$ clusters illustrating the product (dissociated) state.
	Oxygen atoms belonging to the OH$^-$ and H$_3$O$^+$ ions are shown in blue, including the hydrogen-bonded
	molecules in a balls-and-sticks depiction.
		\textbf{b:} Periodic model for the water slab  of width 1 nm.
		\textbf{c:} Free energy profiles computed for the dissociation of water in the bulk phase,
	and for the model systems represented above.
	The coordination number alludes to the proton coordination of the oxygen atom 
	in the donor water molecule.}
  \label{free-energies}
	\end{center}
\end{figure}

Using this strategy,  we computed the barrier to dissociation for isolated
water clusters of 6, 12, and 20 molecules of H$_2$O at 298 K. 
The results are presented in Figure \ref{free-energies},
together with the curve corresponding to bulk water at the same temperature.
Something remarkably surprising happens: the free energy profile for the aggregate of 12 molecules,
with an average diameter not larger than 1 nm, does not differ, within the error of the method, from the
one obtained for the system of 20 molecules, and the two of them are in turn indistiguishable
from the curve corresponding to the bulk phase. Since the dissociation of water creates two ions, 
it can be expected that their solvation energy would be strongly modulated by the size of the 
hydrogen bond network. Instead, a change in the barrier is only observed for the smallest system 
consisting of 6 molecules, where $\Delta G_d$  increases
by nearly 30\%.

\begin{figure}
        \begin{center}
    \includegraphics[width=8.0cm]{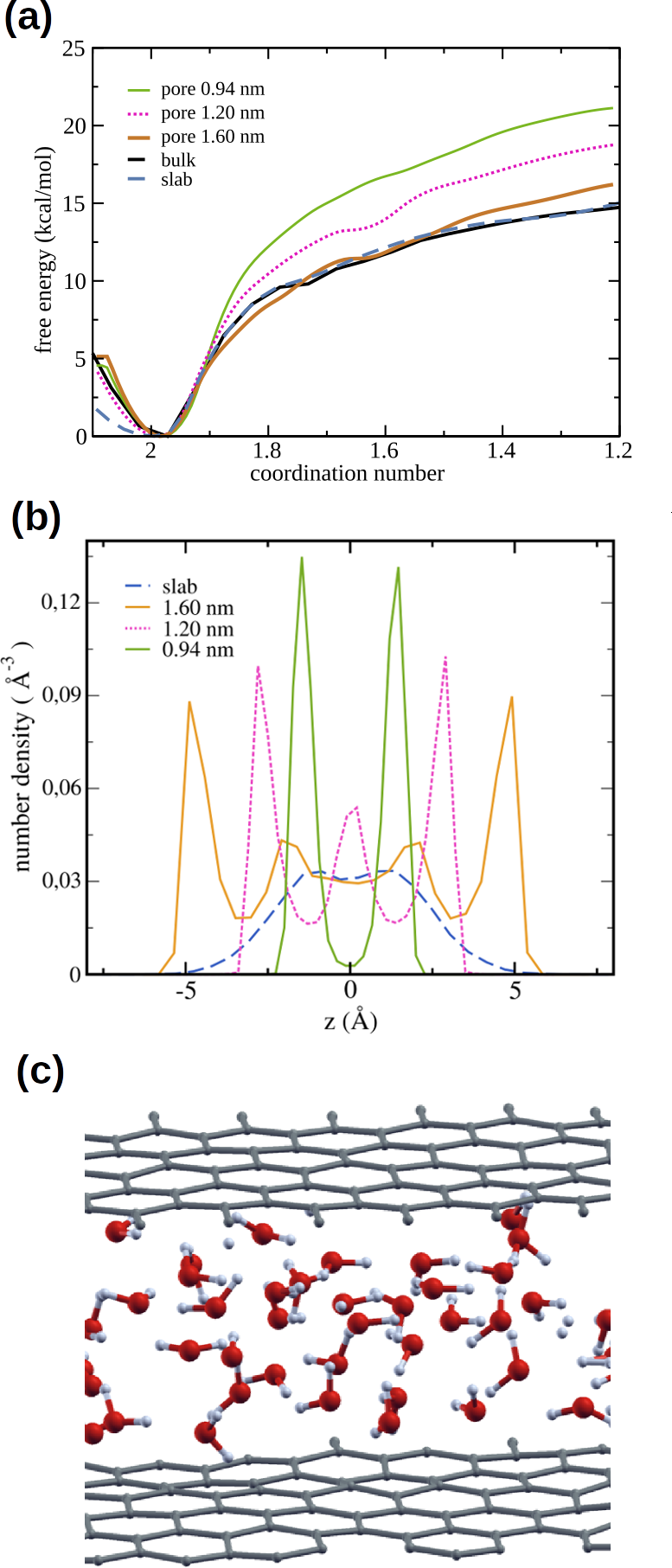}\\

		\caption{\textbf{a:} Free-energy profiles computed for water dissociation within
		slit pores of different widths, delimited by graphene layers of moderate hydrophilicity
		($\theta$=42$^\circ$). The profiles corresponding to bulk water and a 1 nm periodic slab
		in the gas phase are also given for comparison.
		\textbf{b:} Water density profiles as a function of the spatial coordinate 
		perpendicular to the
		principal axis, for the pores and the slab.
		\textbf{c:} Model of the slit pore of graphene, of width 0.94 nm.}
		\label{fig-pores}
        \end{center}
\end{figure}

A similar trend is found for water in slit nanopores. The dissociation
barrier was
studied at 298 K for water confined between graphene layers separated by 0.94, 1.2, and 1.6 nm.
The smallest of these pore models is represented in Figure \ref{fig-pores}c.
We employed for these simulations a multiscale quantum-mechanics
molecular-mechanics (QM-MM) scheme\cite{qe-qmmm,jpcl}
where the graphene 
plates were described with an atomic force-field, and the aqueous phase with DFT
(see details in  Supporting Information).
The use of this treatment has a twofold motivation: whereas it makes affordable the large number of
large-scale simulations needed to complete the analysis, it also offers the possibility of
tuning the water-wall interaction, hence providing a way to assess the effect of
hydrophobicity. We have considered two different interactions that produce
weakly hydrophilic and  moderately hydrophobic contact angles ($\theta$) 
of 42$^\circ$ and 86$^\circ$ respectively.
\cite{jpcb_107_1345} The interlayer space was filled according to the water content
determined from classical molecular dynamics simulations of
graphene pores immersed in an aqueous liquid
reservoir in contact with a gas phase.
The reservoir was thermalized at 298 K and was exposed to a vacuum region to
mimic the air-water interface, thus reproducing ambient conditions
(see Supporting Information).

The free-energy curves corresponding to a water-graphene contact angle of
42$^\circ$, displayed in Figure \ref{fig-pores}a, reveal the same behavior that was
observed for the water aggregates: a spacing of just 1.6 nm is enough
to recover the bulk limit. A very similar result is obtained for
$\theta$=86$^\circ$, suggesting that the effect is inherent to confinement
regardless of the affinity for the interface, at least in the absence of specific interactions 
(see Supporting Information).
The density profiles presented in Figure \ref{fig-pores}b show an
excluded volume involving a fringe
of more than 2 \AA ~adjacent to each graphene plate, leaving in the smallest pores
accessible interlayer spaces of only $\sim$0.5 nm and $\sim$0.7 nm respectively,
comparable to the dimensions of the (H$_2$O)$_6$ cluster.
The effective width in the 1.6 nm pore reaches nearly a nanometer, closer
to the size of the (H$_2$O)$_{12}$ structure, which seems to be sufficient
to restore the bulk dissociation barrier.

This unanticipated outcome, i.e., the insensitivity of water dissociation with respect to system size, 
is consistent with the data from \textit{ab-initio}
simulations presented by Liu and coworkers, showing
that  in 1.5 nm slit pores of  layered neutral clay models, $K_w$ was
essentially the same as in bulk water \cite{geocosmo_75_4978}. 
These authors reported that the acidity was increased in the presence
of interstitial Mg$^{2+}$ cations, while remaining almost unaltered
in a neutral environment, attributing to the charge, and not the
confinement itself, the rise in $K_w$.
On the other hand, this seems to be at odds with the more recent work from
Marx's group \cite{prl_119_056002}, ascribing to bidimensional confinement in slit pores of 1 nm,
an enhancement of 55$\times$ in  the self-dissociation constant $K_w$.
This disagreement  is rather puzzling, given that 
they share the same  methodology, based on
DFT Car-Parrinello molecular dynamics and
the proton coordination $n_H$ as the reaction coordinate.
The origin of this apparent contradiction, however, can be easily tracked to the water content 
implemented in each of these simulations. 
A critical aspect to get meaningful barriers,  contrastable with the bulk values,
is the filling of the nanopores, because the dissociation free energy depends on the pressure, which
is extremely sensitive to the amount of water inside the model pore unit cell.
In the $NVT$ simulations by Mu\~noz-Santiburcio and Marx, the density, 
and not the pressure, was the thermodynamic
variable under control \cite{prl_119_056002}. In particular, the number of water 
molecules $N$ filling the pore was fixed to yield
a target density $\rho$, with $N = \rho \times V_{eff}$, where $V_{eff}$ was arbitrarily chosen.
This resulted in an overcompression of the fluid with respect to the target pressure 
(see the Supporting Information for details) that
produced the observed enhancement in $K_w$  attributed to a confinement effect---the
dissociation constant of water is known to grow as a function of pressure \cite{pkw_vs_pressure}.
In fact, those simulations recovered the bulk value of the pKw when  $\rho$ was reduced by 35\%
(Figure 3 in ref. \cite{prl_119_056002}).
In the study by Liu et al.,\cite{geocosmo_75_4978} instead, or in our own simulations,
the number of water molecules 
included in the supercell was equilibrated with a reservoir at ambient pressure.

To corroborate this interpretation, we applied the same protocol to compute the water dissociation free-energy
in an extended slab of 1 nm width in the vacuum at 298 K.  This slab consisted of 32 H$_2$O
molecules contained in a supercell of lateral dimensions 14.69$\times$14.69 \AA$^2$, 
approximately the same as those
used in ref. \cite{prl_119_056002}.
These simulations recreate the conditions of
bidimensional confinement, avoiding  any compression effects---since the water structure
can equilibrate with the gas phase---and suppressing the interactions with the pore walls.
In this way, the emerging free-energy profile 
expresses the sole effect of 2D confinement.  This profile is depicted in
Figures \ref{free-energies} and \ref{fig-pores}a, where it can 
be seen that $\Delta G_d$ turns out to be---within
the error of our method---not different from the one corresponding to  the bulk liquid.

\begin{figure}
	\begin{center}
    \includegraphics[width=8.6cm]{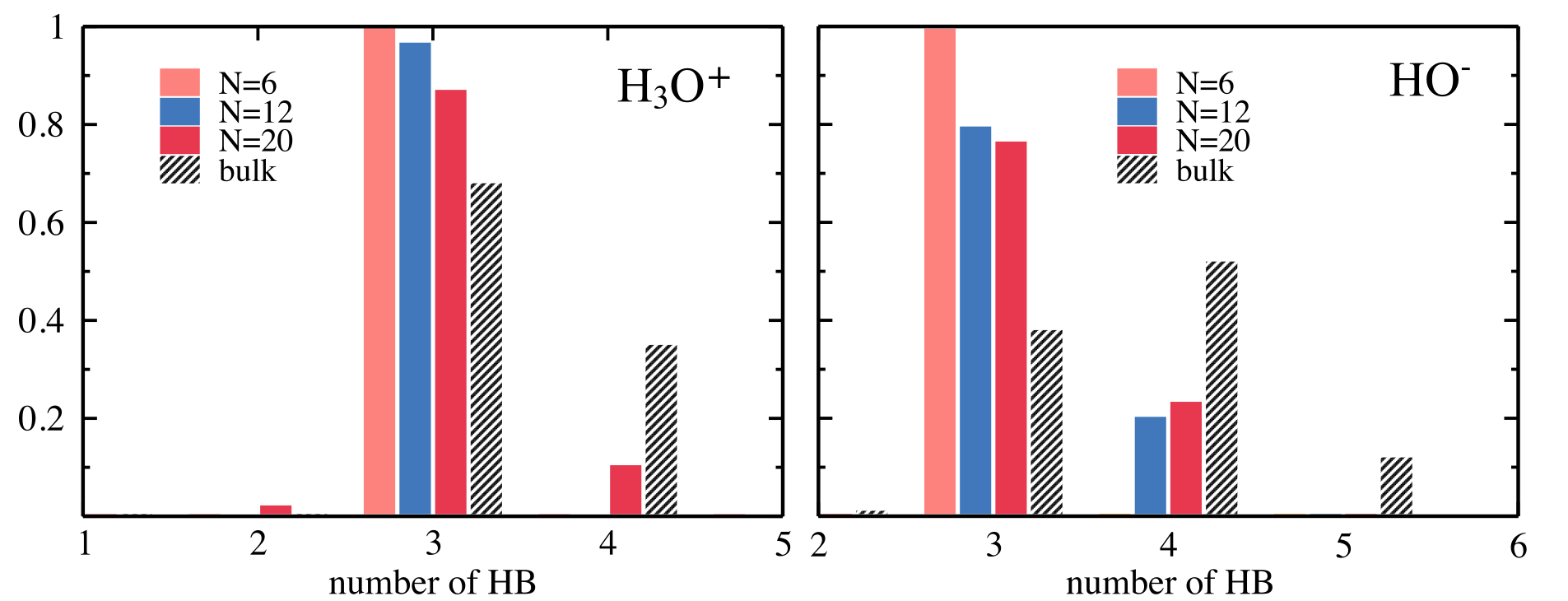}
  \caption{Distribution of the number of hydrogen bonds formed by the hydronium and the hydroxide ions
	in the clusters and in the bulk phase.}
	\label{hbonds}
	\end{center}
\end{figure}

The question that naturally arises is: what is the rationale for the observed trend in the pKw?
A simple explanation can be given in terms of the hydrogen bond structure. The dissociation free-energy
is determined by the thermodynamic stability of the products, 
hydronium and hydroxide, which will in turn depend on the ability of the
hydrogen bond network to solvate these ions. Figure \ref{hbonds} summarizes
the information regarding the number of hydrogen bonds
in the clusters and in the bulk, showing  that
the OH$^-$ anion loses one hydrogen bond in the smallest
cluster with respect to the rest of the environments.
In  the bulk phase,
the hydroxide accepts 4 strong hydrogen bonds and donates a weak one \cite{chemrev-ali}.
A  detailed analysis of the hydrogen bond network, as the one provided in
Table \ref{tablaHB}, reveals that in
the 6 molecules aggregate, and also in the pore of 0.94 nm, the hydroxide accepts only 3 hydrogen bonds,
while it accepts up to 4 in all the other systems.
This missing bond appears to be the reason for its destabilization and the rise in the barrier when $N$=6,
or when the interstitial spacing is 1.20 nm or less.
Moreover, the hydronium ion donates 3 strong and accepts one weak hydrogen bond in the
bulk. In the clusters and in the narrowest pore the latter interaction is lost, however this does not appear 
to significantly  affect the thermodynamics of  dissociation. 
The fact that the oxygen of the hydronium is rather hydrophobic \cite{chemrev-ali}
can explain why placing it in a small cluster  does not incur a large energetic cost.
The hydrogen bond distribution  shows in the smallest pore a more complex behavior
than in the clusters: in the former, the distribution is broader,
with the number of H-bonds varying from 1 to 5.
The net effect is in any case the loss of one of the strong bonds of the OH$^-$ anion,
similarly to what is observed for the cluster of 6 molecules.
In the larger pore, instead, the hydrogen bond network around the ions resembles that of bulk water
(see  Supporting Information for a discussion on the hydrogen bond network).

\begin{table}
	\centering
	\caption{Average distribution of hydrogen bonds for the hydroxide
	and hydronium species.}
\begin{tabular}{cccc}
  &  & donates & accepts \\
\hline
 BULK & H$_3$O$^+$ & 3 & 1 \\
      & OH$^-$     & 1 & 3-4 \\
\hline
 N=20 & H$_3$O$^+$ & 3 & 0 \\
      & OH$^-$     & 0 & 3-4 \\
\hline
 N=12 & H$_3$O$^+$ & 3 & 0 \\
      & OH$^-$     & 0 & 3-4 \\
\hline
 N=6  & H$_3$O$^+$ & 3 & 0 \\
      & OH$^-$     & 0 & 3 \\
\hline
SLAB  & H$_3$O$^+$ & 3 & 0 \\
      & OH$^-$     & 0 &  3-4 \\
\hline
pore 0.94 nm & H$_3$O$^+$ & 3 & 0 \\
      & OH$^-$     & 0 &  3 \\
\hline
pore 1.60 nm & H$_3$O$^+$ & 3-4 & 1 \\
      & OH$^-$     & 1 &  3-4 \\
\hline
\end{tabular}
	\label{tablaHB}
\end{table}

\vspace{0.4cm}
{\bfseries What it takes to drive the ions apart}

\vspace{0.2cm}

Up to now, the rationalization of the dissociation free energy has namely involved the energetic 
contribution of breaking the covalent bond and the reorganization of hydrogen bonding. 
Nevertheless,
once the ions are formed, they must separate away from each other creating solvent screened entities. 
The underlying mechanism by which this happens and the corresponding thermodynamics has been a topic 
of lively discussion in the literature. First-principles simulations by Geissler and co-authors have shown
that the stabilization of the products in the autoionization
of water requires the ions to be separated by at least
three bonds; in that situation, a fluctuation that interrupts the ``hydrogen bond wire'' 
facilitates the diffusion of the proton through the Grotthuss mechanism,
thus  leading to the
effective separation of the hydroxide and hydronium species \cite{parrinello-science}.
More recent work by Hassanali and co-workers has shown the hydrogen bond wire undergoes 
collective compressions which are essential for the proton transfer and 
hydronium-hydroxide separation \cite{HassanaliGibertiCunyKuhneParrinello2013}. 
Using machine-learning approaches, van Erp and co-workers proposed that besides the wire compression, 
factors such as the alignment of the hydrogen bond wire and the extent of tetrahedrality 
of the water molecules also play important roles \cite{MoqadamE4569}.

In the clusters, their own size restrains the maximum separation that
the ions can attain. The 
average hydroxide-hydronium distances ($d_{12}$) 
fluctuate around 2.5-3.5 \AA~during the molecular dynamics sampling 
in the different aggregates (Supporting Information), reflecting
that these ionic species are rarely separated by more 
than one hydrogen bond.
This behavior is also observed in the bulk: the trajectories 
show that even for the lowest
values of $n_H$, the  hydroxide and the
hydronium tend to reside not futher than two bonds apart, and that they may even be in contact
most of the  time. The dimensions of the supercell are in principle large enough to
enable a separation between the counterions of up to five hydrogen bonds, or $d_{12}\sim$8 \AA. 
The fact that distances beyond 4  \AA~are seldom explored suggests
that the use of $n_H$ as the reaction coordinate may not be effective 
to create fully solvated and decorrelated hydronium and hydroxide ions. 
It is thus reasonable to wonder whether our computations are missing
an additional contribution to the free energy associated with a further delocalization
of the ions that may account for a difference between the bulk and the confined systems.

To enforce a full separation of the ions, a new reaction coordinate can be defined
composed of the proton coordination of two different, distant oxygen atoms, $n_H(1)$ and
$n_H(2)$. By setting the reaction coordinate as the difference between these two, 
$\xi(\Delta n_H) = n_H(2) - n_H(1)$, and sampling the phase space from $\xi$=0 to $\xi$=2,
it is  possible to drive a proton from a given water molecule 
(containing oxygen 1) to another chosen molecule (containing oxygen 2) lying far away,
thus producing a well separated OH$^-$-H$_3$O$^+$ pair.
The distance between these two water molecules may slightly 
vary during the process, however, their diffusion is negligible 
within the elapsed time.

\begin{figure}
	\begin{center}
    \includegraphics[width=8.6cm]{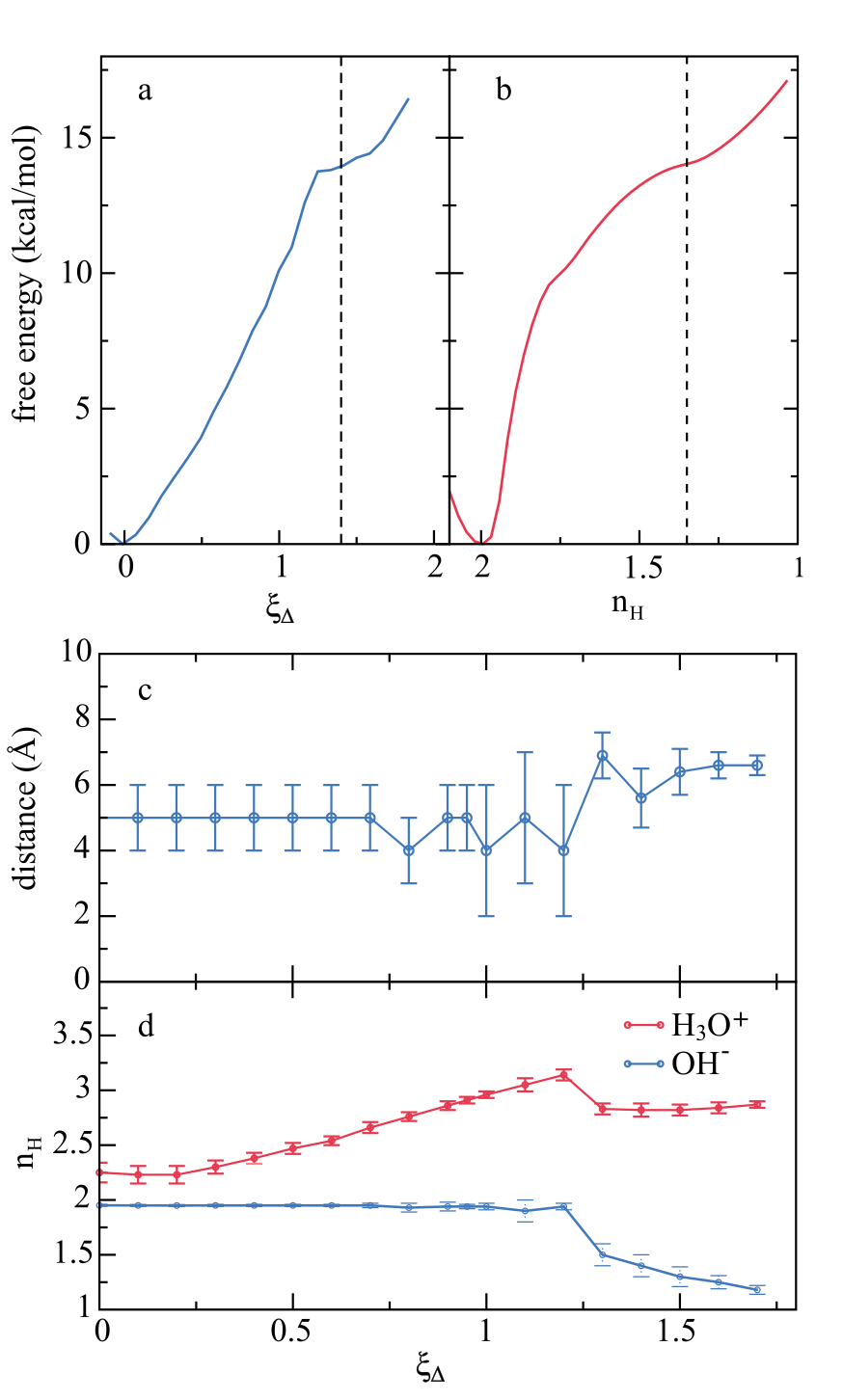}
	\caption{Evolution along the reaction coordinate. 
		\textbf{a} and \textbf{b:} Free energy profiles for the dissociation of water
        in the bulk phase, computed with the single-oxygen reaction coordinate $n_H$
        (red curve) and with the global reaction
        coordinate $\xi(\Delta n_H)$ (blue curve).
        The dashed vertical line marks the reaction endpoint for both coordinates.
		\textbf{c:} Average separation $d_{12}$ between the hydroxide and hydronium species as a function
        of the global reaction coordinate $\xi(\Delta n_H)$.
	This separation is defined and computed as the distance between the oxygen
atoms exhibiting the maximum and minimum proton coordinations. At the early stages of the 
	reaction, when the proton is still bound to the parent water molecule, the 
	fluctuations of the hydrogen-bond network randomly
switch the identity of the minimum coordination oxygen atom, and therefore
the value of $d_{12}$ turns out to be an average of all  O-O distances in the supercell
	($\sim$ 5 \AA). It is only when the reaction is close to completion that $d_{12}$ 
	achieves the targeted value.
		\textbf{d:} Proton coordination number of the oxygen atoms in the
        donor (blue) and acceptor (red) water molecules, as a function of the reaction coordinate.}
  \label{sep-profile}
	\end{center}
\end{figure}

Figure \ref{sep-profile}a presents the free energy  resulting from
this scheme (blue curve) with oxygens 1 and 2 initially separated by $\sim$8 \AA. 
In addition, the  profile obtained earlier is also shown in red on the right panel 
for easy comparison.
To interpret the free-energy curve in blue, it is first important to identify the reaction endpoint.
This can be associated with a value of $n_H$ equal to the coordination
number corresponding to free aqueous OH$^-$ in the bulk phase, which, according to our simulations,
turns out to be around 1.35 \cite{jpcl}.
Figure \ref{sep-profile}c shows that this coordinate effectively draws the hydroxide and
hydronium species apart over the target distance.
In Figure \ref{sep-profile}d,
$n_H(1)$ and $n_H(2)$ are plotted as a function of the global reaction coordinate $\xi$. In particular,
$n_H(1)$ reaches a value of 1.35 when $\xi \sim$ 1.4, implying that at this point the
reaction is complete. In this region the free-energy profile comes to a plateau  
at an energy that corresponds almost exactly to the value of $\Delta G_d$ resulting from the sampling based
on the proton coordination of a single oxygen, shown in the same graph.
The 2D free-energy surfaces recently reported by Joutsuka shed further light on this discussion:
they indicate that, upon dissociation, once the coordination number has reached its final
value of $\sim 1.3$, the free energy turns out to be quite flat for separations extending
between $\sim$2.5 and $\sim$ 7 \AA \cite{joutsuka}.
For the model of 256 molecules,  a small stabilization, in the order of 1-2 kcal/mol,  occurs
when the separation increases to 9 \AA, concertedly with a rearrangement of the hydrogen bond network.
This represents a smooth minimum in the global free-energy landscape which is hard to detect 
in the smaller models. Aside from this, a modest size dependence was found: 
the qualitative features of the potential of mean force
remained unchanged with respect to the system of 64 molecules,
with quantitative differences not larger than 2 kcal/mol \cite{joutsuka}.
It is to be noticed that the value of $n_H$ corresponding to the free-energy minimum identified
in the largest system, of 1.27, 
is essentially the same that gives the inflection point in Figure \ref{sep-profile}a.
The quality of the
free-energy curve produced with our 1D coordinate appears thus comparable to the one resulting
from the 2D potential of mean force, but at a much lower computational cost.
Yet, as a consequence of the supercell size which prevents from reaching separations beyond $\sim$ 7 \AA,
this minimum is unseen in our simulations.
Whether our 1D reaction coordinate $\xi$ is able to reproduce this minimum for
larger sizes will be the subject of future work.

All in all the comparison of the curves in Figures \ref{sep-profile}a and \ref{sep-profile}b,
along with the 2D free-energy maps presented by Joutsuka, 
provide two fundamental insights: (i)
the spatial decorrelation of the nascent H$_3$O$^+$ and OH$^-$ ions contributes a negligible
fraction of the overall free energy, meaning that the barrier is
dominated by the dissociation  step involving breaking the covalent bond that produces the contact ion-pair;
(ii) a complete separation of the ions, attainable either through 2D sampling or through the global reaction
coordinate $\xi$, is needed to reach a minimum or at least a metastable region in the free-energy profile, that
is not observed with the adoption of a reaction coordinate based on a single oxygen atom.


\begin{figure*}
	\begin{center}
    \includegraphics[width=\textwidth]{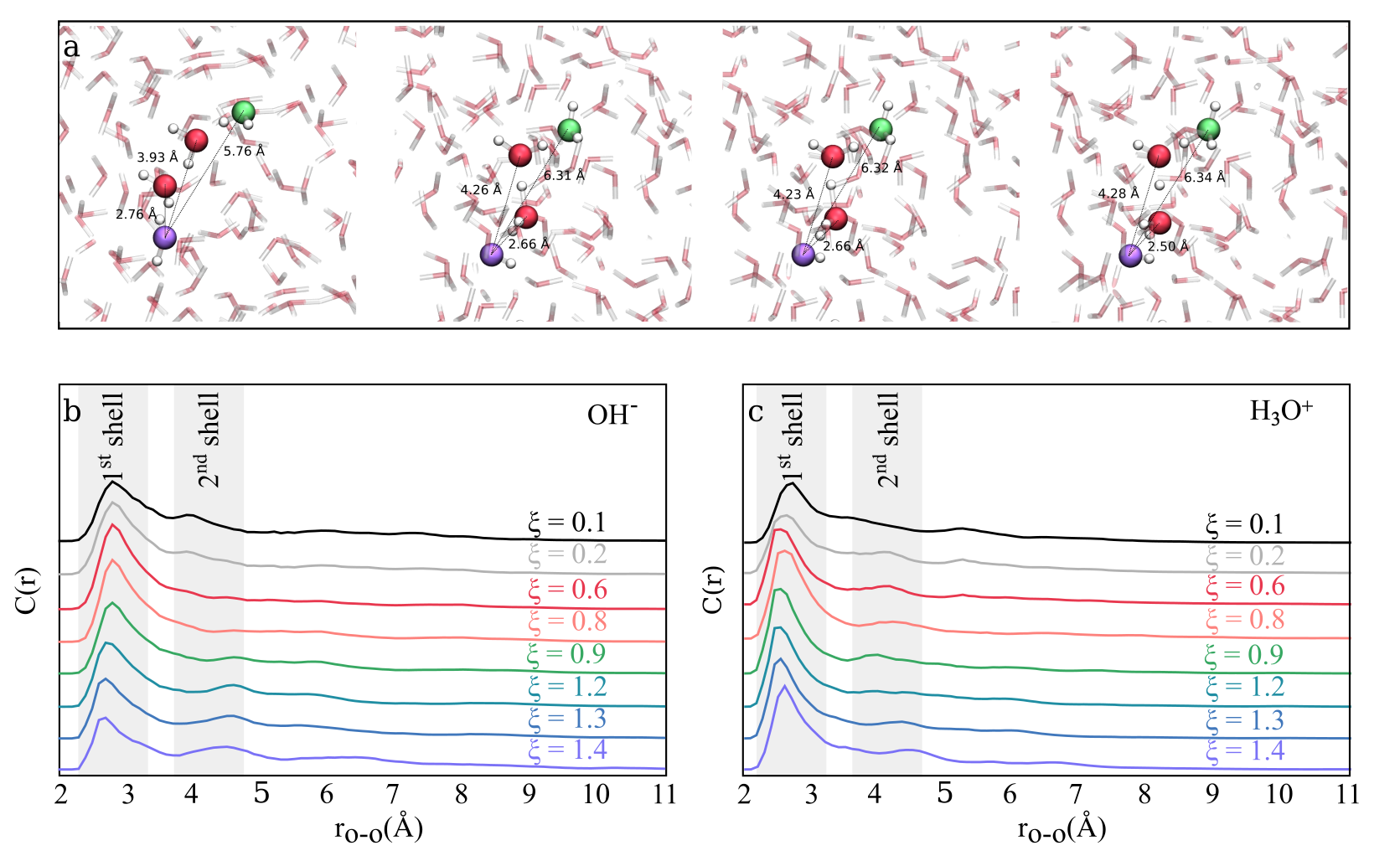}
	\caption{Characterization of the self-dissociation mechanism. 
		\textbf{a:} Snapshots selected from umbrella sampling trajectories in the bulk phase,
	illustrating the dissociation mechanism. Break of the O-H bond leaves a
	contact ion pair, until a proper alignment of the water wire (depicted in indigo)
		allows for a barrierless diffusion  of the proton via the Grotthuss mechanism. \textbf{b:} Fluctuational correlation functions $C(r)$, quantifying the coupling between the dynamics of the nascent OH$^-$ ion and the water molecules placed at $r$, computed for different values of the global reaction coordinate $\xi$. \textbf{c:} Same as \textbf{b} for the H$_3$O$^+$ ion.}
  \label{diss-snapshots}
	\end{center}
\end{figure*}

Further examination of  Figure \ref{sep-profile}d reveals that a gradual protonation
of the acceptor H$_2$O, effected at the expense of the partial dissociation of neighboring molecules,
precedes the deprotonation of the donor. The coordination
number of  the latter, $n_H(1)$, remains approximately 
constant while the system climbs up the free-energy surface, up to $\xi \sim$ 1.2,
when it suddenly drops to its endpoint value. Concurrently with this last event, 
the average separation between the
hydroxide and the hydronium species increases by several angstroms.
The fluctuations that occur within the hydrogen bond network to facilitate 
the sequence of events leading to the separation of the contact-ion pair, 
illustrated in Figure \ref{diss-snapshots}a, have been the subject of several previous 
studies \cite{parrinello-science,HassanaliGibertiCunyKuhneParrinello2013} and are in line
with the results of ref. \cite{joutsuka}.

The contribution of the hydrogen bond network beyond the first solvation shell to the self-dissociation of water can be interpreted in terms of the fluctuational correlation functions $C(r)$ based on the fluctuations of the atomic positions. These correlations quantify the entanglement between the 
dynamics of the incipient ions and that of the water molecules situated at a distance $r$ 
(see Supporting Information). Figures \ref{diss-snapshots}b and \ref{diss-snapshots}c 
evidence a dynamics of the acceptor and donor water molecules primarily correlated with or influenced 
by their first solvation shell. As  the global coordinate evolves towards the formation of the ions, 
the relative influence of solvation beyond the first neighbors decreases substantially, 
reaching a minimum at $\xi \sim 0.6$. At this point, the dynamics of the nascent ions is 
only coupled to their first solvation layer, independent of the fluctuations taking place 
further away. For larger values of $\xi$ until completion of the ionization, the second solvation 
shell slowly regains influence over the ion dynamics (clearly illustrated by the 
relative fluctuational correlation functions presented in the Supporting Information).

The increasing decoupling between the ions and their second solvation shells during the initial stages of the self-dissociation process attest that the dynamics of the reactants will be largely unaltered by perturbations beyond the closest neighbors. This fact is consistent with the insensitivity of the free-energy profile to size reduction beyond the first shell, reinforcing the mechanism proposed above: the formation of the contact ion pair
takes most of the free-energy rise, eventually followed by 
a barrierless proton diffusion propitiated by a favorable alignment of the hydrogen bond network, manifest in the last part of the mechanism ($\xi \sim 1.3$) where $C(r)$ presents some structure beyond the first neighbors.

\section*{Conclusion}
\label{conclusion}
	
Through the application of first-principles biased molecular dynamics simulations combined with an
innovative reaction coordinate, the
present study conclusively establishes that the dissociation free energy of water
is determined by the initial break of the O-H covalent bond. The subsequent separation
of the formed ions, via Grotthuss diffusion, entails only a very marginal fraction of the overall energetic cost.
As a consequence, in the absence of specific interactions at the interface, the pKw
turns out to be practically invariant with respect to system dimensions, providing that they
do not interfer with the solvation of the products.
A sizeable drop in $K_w$ will be only seen when the first solvation
shell of the ions is not intact, and in particular when the
acceptor role of OH$^-$ is affected. In clusters, this occurs at
some point between $N$=6 and $N$=12. In pores exempt from strong interactions, 
the primary solvation structure is destabilized in interstices under 1.6 nm, 
where the acessible space is subnanometric and can hardly accommodate a 
water bilayer or trilayer with molecules exhibiting a hindered motion. In the slab suspended in the gas phase,
of 1 nm width, the translational and
orientational freedom of the molecular dipoles is enough to form a first solvation shell around the ions
that reinstates the bulk dissociation free energy. In fact,
Figure \ref{fig-pores}b  shows that the density profiles for the slab and for the largest pore 
turn out to be very similar in the inner region.
In any case, it is suggestive that the change in the dissociation barrier 
with respect to size stabilizes when the
droplet approaches the length-scale for which hydrogen-bond networks develop, namely, the water hexamer
\cite{science_hexamer2012, science_hexamer2016}.
In this way, dissociation free-energy profiles for aqueous clusters or water in
nanopores  reproduce
the behavior corresponding to the bulk phase, with differences only manifesting
at subnanometer confinement.

In real chemical systems, it is often not possible to consider size effects separately
from interfacial interactions. The presence of charges or polar surface groups,
able to establish specific interactions with H$_2$O and with its ions, will have a disrupting
incidence on aqueous reactivity.
Many examples can be found in the literature discussing the acceleration
or inhibition of reactivity under confinement \cite{angw_55_2, jacs_141_10585, naturenano,
prl_112_028301, nl4008198}.
Leaving aside physical effects as those examined  in
refs. \cite{nl4008198} and \cite{prl_112_028301},
the interactions with the interface appear to be
the main factor controlling
mechanisms and energetic barriers. Indeed,
the present results suggest that when these interactions are not dominant
confinement alone will not affect
the chemistry of water independently of the nature of the interface,
unless system dimensions become comparable to the molecular size.
A broader implication of this finding
is that water's self-ionization constant  at the hydrophobic air-liquid
interface will remain unchanged, warranting further study.


\section*{Acknowledgements}

We are grateful to Dr. Matias Factorovich and
Dr. Mariano Gonzalez Lebrero for useful discussions. This work has been funded by the
Agencia Nacional de
Promoci\'on Cient\'ifica y Tecnol\'ogica de Argentina (PICT 2016-3167 and PICT 2020-02804).
DAS heartily acknowledges the ICTP and its Associates Programme, and the Simons Foundation for support
through grant number 284558FY19.
AH also acknowledges funding by the European Union (ERC, HyBOP, Grant Number: 101043272). Views and opinions expressed are however those of the author(s) only and do not necessarily reflect those of the European Union or the European Research Council. Neither the European Union nor the granting authority can be held responsible for them.

\section*{Conflict of Interest}

The authors declare no conflict of interest.



\bibliography{referencias2}
\bibliographystyle{abbrv}


\end{document}